\documentclass[reprint, superscriptaddress, amsmath, amssymb, aps, pra, nolongbibliography]{revtex4-2}
\usepackage{graphicx,xcolor}
\usepackage{dsfont,braket,siunitx,upgreek}
\definecolor{darkblue}{rgb}{0,0.3,0.7}
\usepackage{hyperref, comment}
\hypersetup{
colorlinks=true,
linkcolor=darkblue,
filecolor=blue,
citecolor=darkblue,  
urlcolor=darkblue,
}

\begin{document}

\preprint{APS/123-QED}

\title{Deployed quantum link characterization via\\Bayesian ancilla-assisted process tomography}
\author{Arefur Rahman}
\email{mrahma75@asu.edu}
\author{Noah~I. Wasserbeck}
\author{Zachary Goisman}
\author{Rhea~P. Fernandes}
\affiliation{School of Electrical, Computer and Energy Engineering and Research Technology Office, Arizona State University, Tempe, Arizona 85287, USA}
\author{Brian~T. Kirby}
\affiliation{DEVCOM US Army Research Laboratory, Adelphi, Maryland 20783, USA} 
\affiliation{Tulane University, New Orleans, Louisiana 70118, USA} 
\author{Muneer Alshowkan}
\affiliation{Quantum Information Science Section, Oak Ridge National Laboratory, Oak Ridge, Tennessee 
37831, USA}
\author{Chris Kurtz}
\affiliation{School of Electrical, Computer and Energy Engineering and Research Technology Office, Arizona State University, Tempe, Arizona 85287, USA}
\author{Joseph~M. Lukens}
\affiliation{School of Electrical, Computer and Energy Engineering and Research Technology Office, Arizona State University, Tempe, Arizona 85287, USA}
\affiliation{Quantum Information Science Section, Oak Ridge National Laboratory, Oak Ridge, Tennessee 
37831, USA}

\date{\today}

\begin{abstract}
The development of large-scale quantum networks requires reliable quantum channels, the quality of which can be quantified by the framework of quantum process tomography. In this work, we leverage ancilla-assisted process tomography and Bayesian inference to probe a 1.6~km deployed fiber-optic link. We send one of two polarization-entangled photons from Alice in one building to Bob in another, exploiting the local qubit as an ancilla system to characterize the corresponding quantum channel. Monitoring over a 24~h period returns a steady process fidelity of 95.1(1)\%, while controllable spectral filtering with passbands from 0.025--4.38~THz finds fidelities that first increase, then level off with bandwidth, suggesting both stable operation with time and minimal polarization mode dispersion. To our knowledge, these results represent the first AAPT of a deployed quantum link, revealing a valuable tool for \emph{in situ} analysis of entanglement-based quantum networks.
\end{abstract}

\maketitle

In the context of quantum networks~\cite{wehner2018}, photons represent ideal information carriers due to their ability to travel long distances in optical fiber with minimal interactions with their environment. Entangled photons enable nonlocal interactions between parties for communication \cite{ekert1991,bennett1992}, distributed quantum computing \cite{monroe2014}, and distributed quantum sensing \cite{eldredge2018,proctor2018}. Amid this wide-ranging field, broadband polarization-entangled photon sources offer valuable near-term possibilities for quantum communications within the telecom infrastructure~\cite{Poppe2004, Treiber2009, wengerowsky2019, wengerowsky2020, Alshowkan2021, Alshowkan_architecture2022, Neumann2022}.

The variation of polarization transformations in optical fiber due to environmental effects has prompted a  string of solutions designed to monitor and correct for time-varying quantum channels, often via classical reference fields that are wavelength- or time-multiplexed with the quantum signals~\cite{Xavier2008, Xavier2009, Treiber2009, peranic2023study, Craddock2024}. Recently, full quantum process tomography (QPT) has been realized over deployed fiber links as well, based on measurements with classical polarization references~\cite{Kucera2024} or dedicated weak coherent states~\cite{Chapman2023}. In this work, we experimentally implement an alternative, \emph{in situ} method for channel characterization over a 1.6~km deployed fiber link on the Arizona State University (ASU) campus. Leveraging a polarization-entangled photon source in which one photon is detected locally and the other transmitted to a second building, Bayesian ancilla-assisted process tomography (AAPT) estimates the complete quantum channel, confirming a process fidelity of 95.1(1)\% and stability over a 24~h period. Spectrally resolved measurements from 25~GHz to 4.38~THz likewise show consistent performance over a wide range of bandwidths. In total, our results forge a path for comprehensive quantum network channel characterization with in-place entanglement resources.

\begin{figure*}[bt!]
     \centering
     \includegraphics[width=\textwidth]{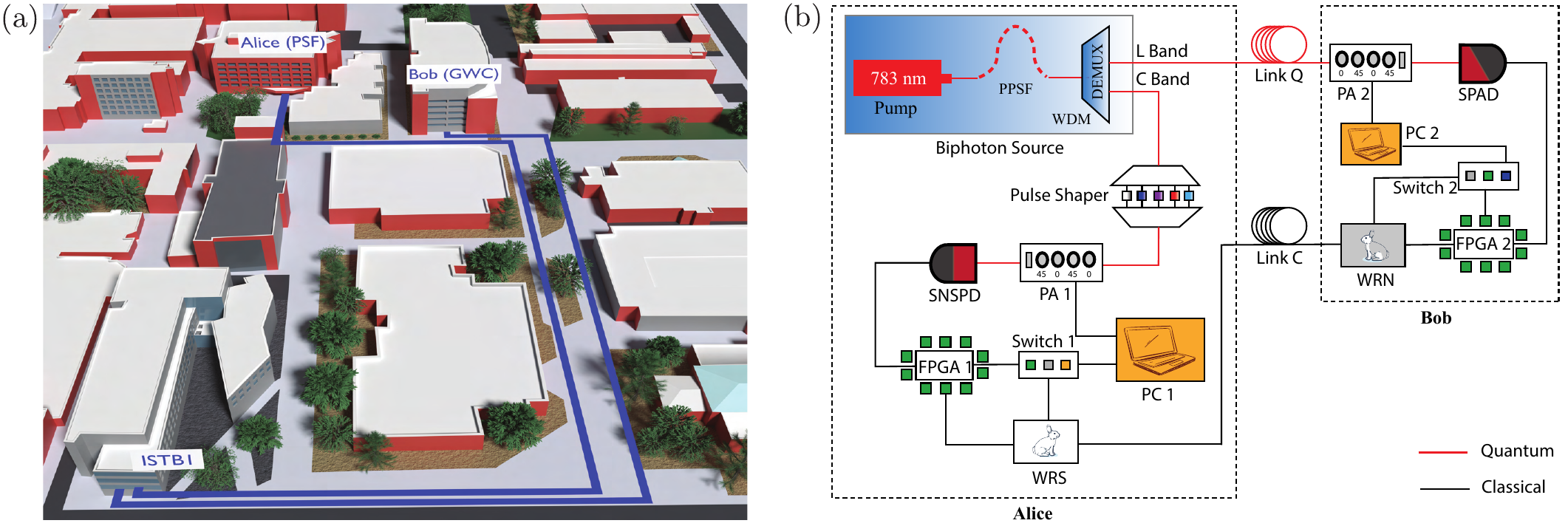}
     \caption{(a) Map of the 1.6~km fiber link on the ASU campus. The blue line extending from Alice to Bob shows the fiber path. (b)~Schematic of the overall setup including devices on both sides of the quantum link. Details are provided in the text.}
     \label{fig1:map_setup}
\end{figure*}

Standard QPT (SQPT)~\cite{Chuang1997,Poyatos1997,Bongioanni2010} involves preparing ensembles of linearly independent quantum states, each of which is subjected to the quantum process and then estimated via quantum state tomography (QST) \cite{nielsen2010}. In contrast, AAPT~\cite{D'Ariano2001,D"ur2001,Altepeter2003} introduces an auxiliary system and prepares the combined system in a single input state such that complete information about the dynamics can be extracted from the state alone. By performing QST of the combined output system we can infer the process acting on the system of interest. Although equivalent to SQPT in terms of informational complexity, AAPT can often offer practical advantages. For example, if an entangled input state is already available---as in the case of an entanglement distribution network---an unknown quantum channel can be characterized automatically, with no modifications to the state preparation or measurement systems.

For concreteness, consider two $d$-dimensional quantum subsystems $A$ and $B$. Both are produced at Alice; $A$ is always detected there, whereas $B$ starts at Alice (for the input) and is transmitted to Bob (for the output). Mathematically, the quantum link to Bob is described by a completely positive trace-preserving map $\mathcal{E}$ defined by
\begin{align}\label{eq1:cptp_map}
    \mathcal{E}(\rho_B) = \sum_{k=1}^R A_k \rho_B A^{\dagger}_k= \sum_{i=1}^{d^2}\sum_{j=1}^{d^2} \chi_{ij} E_i \rho_B E^{\dagger}_j,
\end{align}
where $\rho_B$ is any arbitrary single-qudit state, $R \le d^2$ defines the Choi rank, and $\{A_k\}$ are Kraus operators such that $\sum_k A^{\dagger}_k A_k = \mathds{1}$. As exploited by the second equality in Eq.~(\ref{eq1:cptp_map}), each Kraus operator can be decomposed as $A_k = \sum_{i} c_{ki} E_i$, where $\{E_i\}$ is a set of orthogonal basis operators, $\{c_{ki}\}$ is a set of expansion coefficients, and $\chi_{ij}=\sum_k c_{ki}c^{*}_{kj}$ defines a process matrix expressing the map $\mathcal{E}$ in the $\{E_i\}$ basis.

Complementing the Kraus operator picture, the Choi--Jamiolkowski isomorphism~\cite{Gilchrist2005} 
defines a one-to-one correspondence between the map $\mathcal{E}$ and the joint density matrix $\Phi_\mathcal{E}$ of ancilla system $A$ and primary system $B$ resulting from the specific evolution
\begin{align}\label{eq3:isomorphism}
    \Phi_\mathcal{E} = (\mathds{1}_A \otimes \mathcal{E}_B) (\ket{\phi^{+}}\bra{\phi^{+}}) = \sum_{l=1}^{d^2} e_l \ket{\gamma_l} \bra{\gamma_l},
\end{align}
where $\ket{\phi^{+}}=\frac{1}{\sqrt{d}} \sum_{n=1}^{d}\ket{n_A n_B}$ is the joint input state, $\{\ket{n_A n_B}\}$ comprise an orthonormal basis in the joint state space, and the notation $\mathds{1}_A \otimes \mathcal{E}_B$ signifies the operation of the identity on system $A$ and the process $\mathcal{E}$ on system $B$. Since $\mathcal{E}$ is completely positive, the Choi matrix $\Phi_\mathcal{E} \geq 0$, and hence the spectral decomposition expressed by the second equality in Eq.~(\ref{eq3:isomorphism}) has positive eigenvalues $e_l \geq 0$. 
This eigendecomposition provides a recipe to obtain a valid set of Kraus operators from the Choi matrix $\Phi_\mathcal{E}$, which in our convention can be realized by setting $\braket{i|A_{k}|j} = \sqrt{de_k}\braket{ji|\gamma_k}$~\cite{wood2015}. We apply this freely below to convert between the Choi and process matrix descriptions.

Intuitively, the isomorphism between \emph{process} $\mathcal{E}$ [Eq.~(\ref{eq1:cptp_map})] and \emph{state} $\Phi_\mathcal{E}$ [Eq.~(\ref{eq3:isomorphism})] can be viewed as the justification for AAPT: experimentally finding the quantum map $\mathcal{E}$ on system $B$ is equivalent to estimating the state $\Phi_\mathcal{E}$ in the joint $AB$ Hilbert space. In practice this analogy is not absolute, though, as AAPT does not require a maximally entangled input as assumed in the definition of $\Phi_\mathcal{E}$~\cite{Altepeter2003}.
Any experiment ultimately probes the output through the probabilities of specific outcomes, which for our tests can be taken as projective measurements onto pure states $\{\ket{\varphi_s}\}$. Thus, for the joint input state $\rho_{AB}$ and corresponding output $\tilde{\rho}_{AB}$ the probability to measure outcome $s$ is
\begin{equation}
\label{eq:probability}
p_s = \braket{\varphi_s|\tilde{\rho}_{AB}|\varphi_s} = \braket{\varphi_s|(\mathds{1}_A \otimes \mathcal{E}_B) (\rho_{AB})|\varphi_s}.
\end{equation}
Therefore, with some collection of observed events $\{N_s\}$, one can  infer either the output state $\tilde{\rho}_{AB}$ or (if the input $\rho_{AB}$ is known) the process $\mathcal{E}$.
Historically, three major approaches to estimate a quantum state or process  from measurements are least-squares inversion~\cite{nielsen2010}, maximum likelihood estimation~\cite{James2001}, and Bayesian inference~\cite{Blume-Kohout2010}. Due to its natural uncertainty quantification, optimality in terms of mean squared error, and avoidance of unjustified low-rank estimators, we calculate all quantities of interest---input $\rho_{AB}$, output $\tilde{\rho}_{AB}$, and process $\mathcal{E}$---using Bayesian inference in this work.

At a high level, our method involves parametrizing either the joint density matrix or single-qubit Kraus operators such that they satisfy physicality conditions (e.g., positivity and normalization), defining a prior distribution for the parameters, computing a likelihood based on Eq.~(\ref{eq:probability}) and the results $\{N_s\}$, and sampling from the posterior distribution (product of prior and likelihood) via preconditioned Crank--Nicolson (pCN) Markov chain Monte Carlo~\cite{Lukens2020}. Following \cite{Lu2022b, Chapman2023}, we assume a Poissoninan likelihood for all examples, while the prior depends on the unknown: we assume a Bures prior~\cite{Osipov2010} when estimating either $\rho_{AB}$ or $\tilde{\rho}_{AB}$; when estimating the channel $\mathcal{E}$, we assume a uniform Lebesgue prior on $\Phi_\mathcal{E}$~\cite{Kukulski2021}. We save $2^{10}$ thinned samples from pCN chains of total length $2^{22}$, from which we estimate the mean and standard deviation of any quantity of interest.

We apply Bayesian AAPT to quantum channel characterization on a deployed quantum link on the ASU campus. Figure~\ref{fig1:map_setup}(a) shows the lightpath from Alice's lab in Physical Sciences F (PSF) to Bob's lab in the Goldwater Center (GWC) via a fiber hub in Interdisciplinary Science and Technology I (ISTB1). The $1.6$~km-long link comprises mostly underground optical fibers with one patch station at ISTB1.
Figure~\ref{fig1:map_setup}(b) provides a schematic of the experimental setup. To generate polarization-entangled photon pairs, we use a biphoton source (OZ Optics EPS-1000) based on type-II spontaneous parametric down conversion in periodically poled silica fiber (PPSF)~\cite{Zhu2012}. The wavelength of the pump laser is $783.65$~nm, and the the signal and idler photons fill the C  and L telecom bands, respectively, separated by a demultiplexer centered at $1567.3$~nm (191.28~THz).
Two single-mode optical fibers, denoted as Link Q and Link C in Fig.~\ref{fig1:map_setup}(b), carry the quantum and classical signals, respectively, from Alice to Bob. To explore bandwidth-dependent effects in the channel, a commercial pulse shaper (II-VI WaveShaper $4000$A) with acceptance band $1525.97$--$1568.77$~nm ($191.10$--$196.46$~THz) selectively varies the biphoton bandwidth  through a C-band programmable filter on Alice's local photon.

Alice and Bob measure their photons using polarization analyzers (PA$1$ and PA$2$;  Nucrypt PA-$1000$) which consist of four variable waveplates and one polarizer. The four independently controlled waveplates of the PA-1000 are nominally oriented at $\ang{0}$, $\ang{45}$, $\ang{0}$, and $\ang{45}$ with respect to the polarizer and set the qubit projections for the rectilinear $\{\ket{H},\ket{V}\}$, diagonal $\{\ket{D},\ket{A}\}$, and circular $\{\ket{R},\ket{L}\}$ bases~\cite{James2001}
via personal computers (PC$1$ and PC2). We consider all 36 possible joint projections, so that $\ket{\varphi_s} = \ket{\alpha \beta}$ for $\alpha,\beta \in \{H,V,D,A,R,L\}$ in Eq.~(\ref{eq:probability}).

\begin{figure}[tb!]
     \centering    \includegraphics[width=\linewidth]{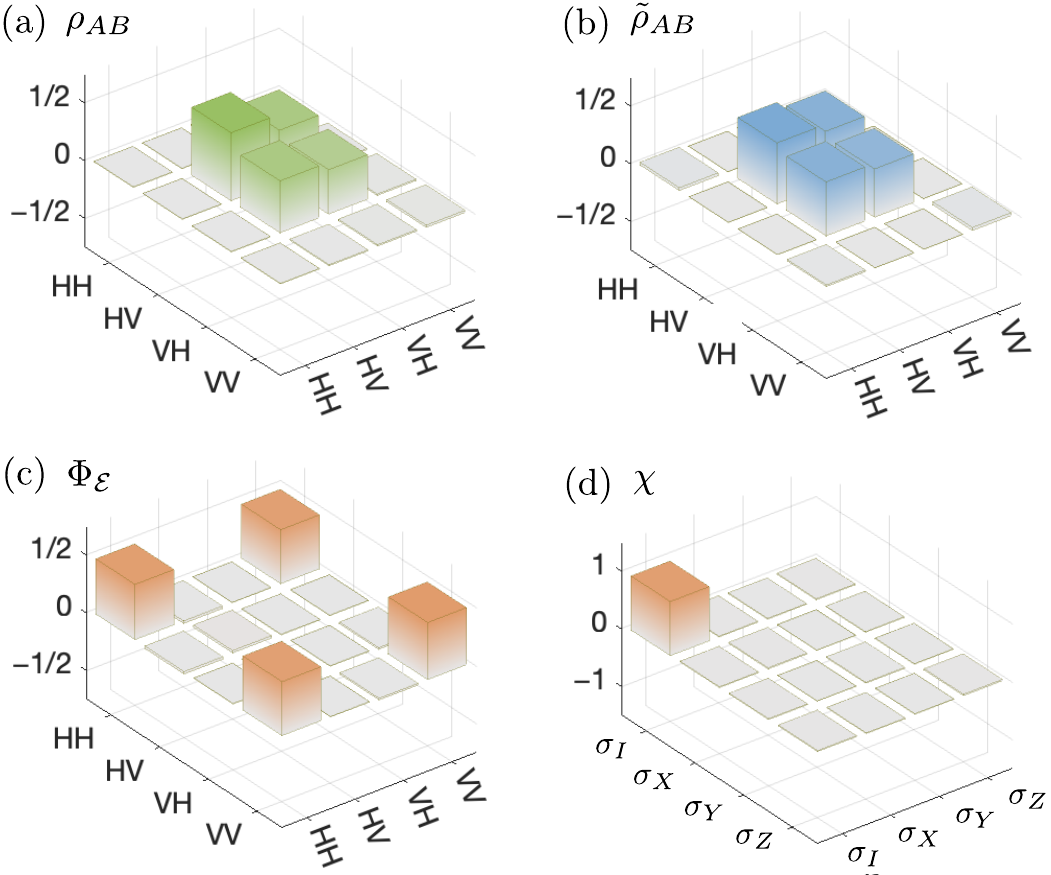} 
     \caption{Real parts of the inferred (a) input state $\rho_{AB}$, (b) output state $\tilde{\rho}_{AB}$, (c) Choi matrix $\Phi_\mathcal{E}$, and (d) process matrix $\chi$, where both $\Phi_\mathcal{E}$ and $\chi$ describe the dynamics of the quantum link. Local rotations have been applied to align reference frames, and all imaginary components (not shown) are less than $0.02$.}
     \label{fig2: matrices}
\end{figure}

System $B$'s qubit is detected by a free-running single-photon avalanche diode (SPAD; IDQuantique) at 25\% detection efficiency and $10$~$\upmu$s dead time. The ancilla qubit $A$ is detected with a superconducting nanowire single-photon detector (SNSPD; Quantum Opus) with $\geq 80$\% efficiency and $50$~ns dead time. The coincidences for each projection are collected over 30~s with a 1.86~ns coincidence window. Following~\cite{Alshowkan_architecture2022}, FPGA development boards (FPGA$1$ and FPGA$2$) are locked to White Rabbit~\cite{Lipinski2011} for time tagging. The White Rabbit switch (WRS) at Alice is connected to the White Rabbit node (WRN) at Bob over the classical fiber link (Link C). Ethernet signals between the switches at Alice and Bob piggyback on the White Rabbit communications, enabling control of all instruments at Bob from PC$1$ at Alice.

We first measure the prepared quantum state $\rho_{AB}$ locally (both photons at Alice), which ideally should be of the from $\ket{\psi^{+}} = (\ket{HV}+\ket{VH})/\sqrt{2}$. With the pulse shaper set to its full acceptance band, we find the Bayesian mean density matrix in Fig.~\ref{fig2: matrices}(a). The fidelity with respect to the ideal state is $\mathcal{F}_\rho = \braket{\psi^+|\rho_{AB}|\psi^+} = 93.5(2)\%$. (In these and all results below, local unitaries have been applied to orient Alice's and Bob's reference frames for maximum fidelity~\cite{Alshowkan2022c}.) Then we send the L-band  photon to Bob through the quantum link.
The produced output  $\tilde{\rho}_{AB}$ [Fig.~\ref{fig2: matrices}(b)] matches the Bell state $\ket{\psi^{+}}$ with $95.4(2)\%$ fidelity---slightly higher than the input, perhaps due to variations from the fiber reconnections and small differences in the specific devices in the local and nonlocal detection setups. 

\begin{figure}[b!]
\centering
\includegraphics[width=\linewidth]{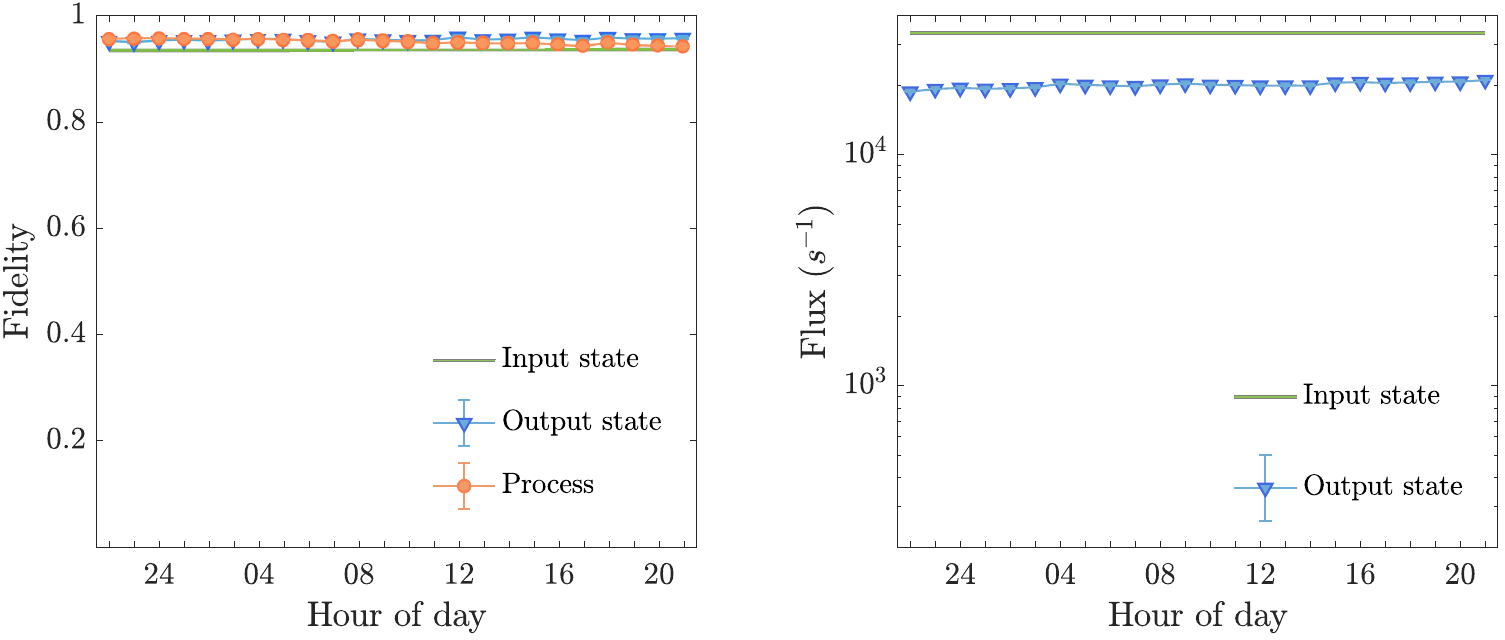}
\caption{(a) State fidelity of the output $\tilde{\rho}_{AB}$ (blue triangles) and  process fidelity of the link $\Phi_\mathcal{E}$ (orange circles) over a 24~h period. (b) Detected photon pair flux of the output state (blue triangles) over the same 24~h period. The fidelity and flux corresponding to the input state are shown in green, with thickness denoting the Bayesian error bars.}
\label{fig2: time_analysis}
\end{figure}

Next, leveraging the same dataset used to infer $\tilde{\rho}_{AB}$, but now considering the AAPT picture [second equality in Eq.~(\ref{eq:probability})] with $\rho_{AB}$ given by the previously determined Bayesian mean [Fig.~\ref{fig2: matrices}(a)], we characterize the channel directly, obtaining the Bayesian mean Choi matrix $\Phi_\mathcal{E}$ in Fig.~\ref{fig2: matrices}(c). In contrast to the states which ideally appear as $\ket{\psi^+}$ Bell states with our type-II source, $\Phi_\mathcal{E}$ should ideally equal the Bell state $\ket{\phi^+}=(\ket{HH}+\ket{VV})/\sqrt{2}$ for the identity channel, [cf. Eq.~(\ref{eq3:isomorphism})], which we confirm via the process fidelity $\mathcal{F}_\mathcal{E} = \braket{\phi^+|\Phi_\mathcal{E}|\phi^+} = 95.1(1)\%$. As an additional, arguably more intuitive depiction of the channel, the process matrix $\chi$ in the Pauli basis [Fig.~\ref{fig2: matrices}(d)] reveals the expected identity, with the only appreciable nonzero element on the term corresponding to $E_i=E_j=\sigma_I$ in Eq.~(\ref{eq1:cptp_map}).
As an aside, we note that although the state ($\mathcal{F}_\rho$) and process ($\mathcal{F}_\mathcal{E}$) fidelity formulas are almost identical---they both compare the matrix of interest with an ideal Bell state---they nonetheless quantify distinct physical entities: the former describes the closeness of a specific \emph{two-qubit} state to a Bell state, while the latter describes how well a channel preserves a \emph{single-qubit} state passing through it.

Stability over time is a critical feature of any quantum channel, defining the timescale over which quantum communications can be performed or the speed with which any compensation system must operate. 
Figure~\ref{fig2: time_analysis} shows the process fidelity and photon flux of the quantum link measured hourly over a $24$~h period beginning at 10~pm local time on 21 September 2024. Although we do apply a numerically chosen local rotation to the first datapoint to align reference frames, we keep this rotation fixed for all other results in Fig.~\ref{fig2: time_analysis} in order to quantify fluctuations. An average process fidelity of $95.1\%$ with an average standard deviation of $0.1\%$ confirms that the channel is highly stable, attributable to the primarily underground fiber lightpath. An average coincidence rate of 20,056 s$^{-1}$ is detected (over all states in a basis) with an average standard deviation of $51$~s$^{-1}$, which compared to the local flux of 33,854~s$^{-1}$  suggests a total channel loss of 2.27~dB.

\begin{figure}[tb!]
\centering
\includegraphics[width=\linewidth]{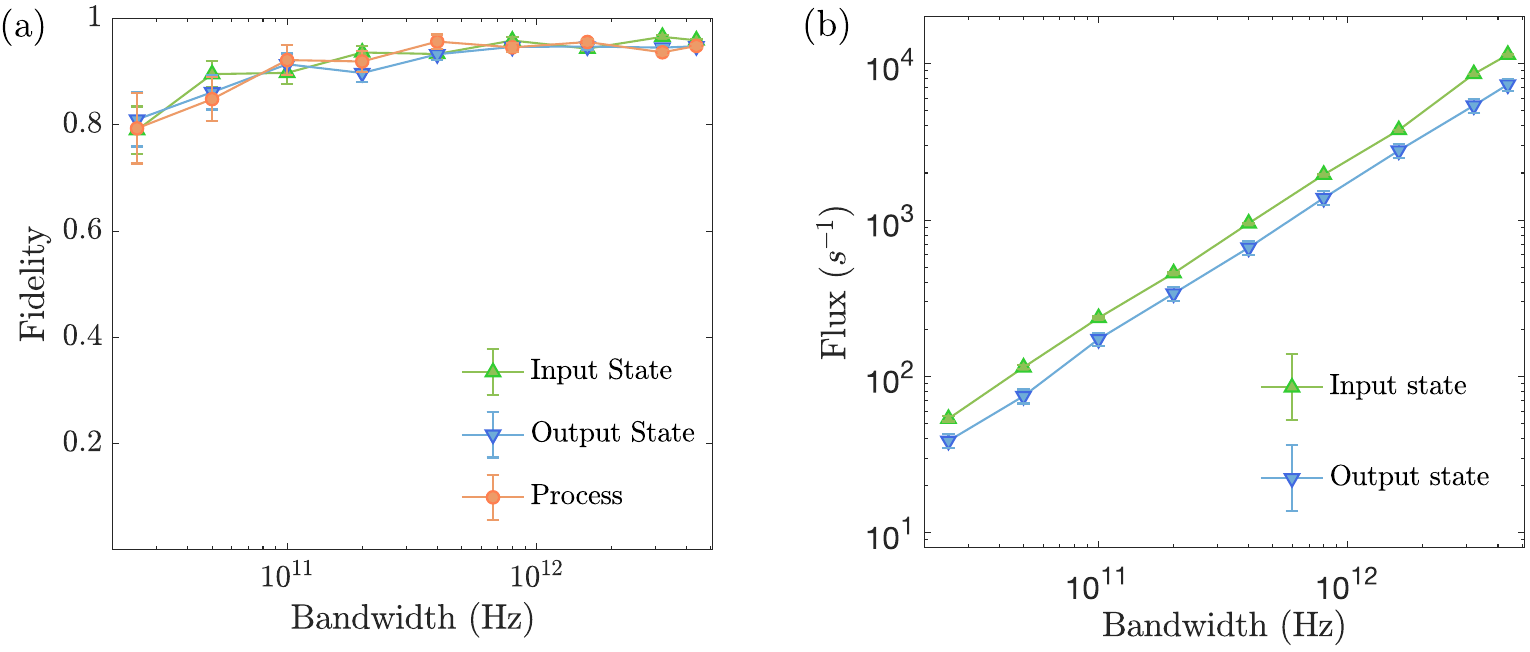}
\caption{(a) State fidelity of the input state (green triangles), state fidelity of the output state (blue triangles), and process fidelity of the link (orange circles) over filter bandwidths from $0.025$ THz to $4.38$ THz. (b) Detected flux corresponding to the input and output states over the same bandwidths.}
\label{fig3_frequency analysis}
\end{figure}

As a second test, we explore the effect of optical bandwidth on the quantum channel fidelity. Although temporal spreading due to chromatic dispersion is expected to have minimal impact over the measured coincidence window of 1.86~ns, polarization mode dispersion (PMD) can lead to strong bandwidth-dependent effects, decreasing fidelity when the polarization-dependent delays are comparable to the inverse photon bandwidth~\cite{Antonelli2011, Brodsky2011}. To tune the effective link bandwidth, we program a bandpass filter on the C-band pulse shaper at Alice,  postselecting the same bandwidth on Bob's photon due to frequency entanglement. Figure~\ref{fig3_frequency analysis} plots the Bayesian-inferred fidelities and flux for filters centered at 193.75~THz with bandwidths increasing from 25~GHz to 4.38~THz---corresponding to the full C-band (1530--1565~nm).

The fidelity actually increases with bandwidth, suggesting not only negligible degradation from PMD, but also dominant noise sources that are flux-based, i.e., caused by background counts whose impact is felt more strongly at lower photon rates. Incidentally, this behavior highlights an advantage of AAPT compared to SQPT based on auxiliary classical lasers which---while experiencing the same polarization transformations as single photons---are not as sensitive to noise sources like Raman scattering, crosstalk, or dark counts that might impact lower flux beams. By characterizing the channel with the entangled resource \emph{in situ}, AAPT automatically includes flux-dependent effects at the level of operation.

Moving forward, it would be interesting to apply our technique to links with significant polarization-dependent loss (requiring non-trace-preserving maps~\cite{Bongioanni2010}) or links where the quantum channel evolves rapidly in time. In the latter context, AAPT could be enlisted to validate a separate polarization compensation system or even incorporated into the feedback loop itself. Extending the method to multiple receivers could bring advantages in multinode network characterization, where recent theoretical efforts in quantum network tomography have targeted specific impairments like bit flips and depolarizing noise~\cite{deAndrade2024}, rather than the fully arbitrary quantum channel attacked by traditional AAPT. Accordingly, the degree to which realistic channel assumptions can simplify AAPT appears a profitable research direction, both in increasing the scalability of AAPT and in bringing network tomography toward experimental realization.

\begin{acknowledgments}
We thank L. Qian for valuable discussions on the entangled-photon source. Funding was provided by Cisco Research and the U.S. Department of Energy (ERKJ432, DE-SC0024257).
\end{acknowledgments}


%

\end{document}